\documentclass[11pt]{article}
\begin{document}
\begin{flushright}
Preprint IC/2002/101 \\ 
hep-th/0208110 \\ 
v2. February 14, 2003. \\ 
To appear in Phys. Lett. {\bf B} 
\end{flushright}

\begin{center}
{\Large
BPS preons and tensionless super-p-branes in generalized superspace} 

\bigskip 

{\bf Igor A. Bandos} 

\bigskip 

{\it Departamento de F\'{\i}sica Te\'orica and IFIC,} 
{\it  46100-Burjassot (Valencia), Spain}

{\it Institute for Theoretical Physics, NSC KIPT, 
UA61108, Kharkov, Ukraine} 

{\it The Abdus Salam ICTP, Trieste, Italy} 

\medskip

\end{center}

\begin{abstract} 
\noindent
Tensionless super--p--branes in a generalized superspace with 
additional tensorial central charge coordinates may provide 
an extended object model for BPS preons, {\it i.e.} 
for the hypothetical constituents of M--theory preserving 31 of 32 
supersymmetries 
\cite{BPS01}.

\end{abstract}

\section{Introduction} 
 
Recently, a new wave of interest for higher spin theories and their 
supersymmetric extensions can be witnessed 
\cite{V01,V01s,Whs,V01c,Bars01,ESPS02}.
Moreover, the study of \cite{BLS99,V01s,V01c} exhibits the relation 
among massless high--spin theories and simple particle--like 
dynamical models 
\cite{BL98,BL98'} living in generalized superspace 
$\Sigma^{({n(n+1)\over 2}|n)}$ with local coordinates  
\begin{eqnarray}\label{ZM} 
Z^{{\cal M}}=(X^{\alpha\beta}\, , \; \theta^\alpha )\; , 
\quad X^{\alpha\beta}= X^{\beta\alpha}\; ,  
\quad \alpha, \beta = 1, \ldots , n \; , 
\end{eqnarray}
\cite{HP82C88,JdA00}.  
This relation suggested 
a way to introduce a concept of causality in 'symplectic spacetime' 
$\Sigma^{({n(n+1)\over 2}|0)}$ \cite{V01c} 
({\it i.e.} in the bosonic body of $\Sigma^{({n(n+1)\over 2}|n)}$) 
parametrized by  symmetric $GL(n)$--tensor coordinates 
$X^{\alpha\beta}=X^{\beta\alpha}$ \cite{Fr86}.

For $n=2^k$, where $\alpha$ can be treated also as a spinor index 
of a $D$--dimensional 
Lorentz group $SO(t,D-t)$ for some $D$ and $t$,  
$X^{\alpha\beta}=X^{\beta\alpha}$ 
can be regarded as symmetric spin--tensor coordinates.  
For  $k>1$ the set of such bosonic coordinates includes, 
besides the usual $D$--dimensional spacetime 
coordinates $x^\mu = X^{\alpha\beta} \Gamma^\mu_{\alpha\beta}$, a set of
antisymmetric tensorial coordinates 
$y^{\mu_1 \ldots \mu_q}= X^{\alpha\beta} 
\Gamma^{\mu_1 \ldots \mu_q}_{(\alpha\beta )}$ 
($y^{\mu \nu}$, $y^{\mu_1\ldots \mu_5}$ for $D=11$ generalized  superspace 
$\Sigma^{(528|32)}$). Just the introduction of gamma--matrices or, 
equivalently, the distinction between vector and antisymmetric tensor 
coordinates  breaks the manifest $GL(n)$ symmetry of the generalized 
superspace down to $Spin(t,D-t)$.  
(Note that $n=32$ case allows also $SO(2,10)$ interpretation, in which 
$X^{\alpha\beta}$ contains antisymmetric tensor coordinates only 
\cite{Bars,RS,Manvelyan}).   
Such a breaking of high spin $GL(n)$ symmetry (actually of the 
$OSp(2n|1)$ symmetry, 
see \cite{V01,V01s,V01c,BL98,BLS99} and 
Sec. 4 below) 
is expected to be spontaneous. 

An important property of the models \cite{BL98,BL98'}, 
not yet 
reflected in higher spin theories, is that they 
describe BPS states preserving all but one spacetime supersymmetries. 
This property is closely related to the fact 
that these models produce the generalized Penrose relation 
\begin{equation}\label{Pll}
P_{\alpha\beta} = \lambda_{\alpha} \lambda_{\beta} 
\end{equation} 
({\it cf.} \cite{Pen}) 
as a constraint for the momentum $P_{\alpha\beta}(\tau)$ canonically 
conjugate to the coordinate function $X^{\alpha\beta}(\tau)$, 
\begin{eqnarray}\label{Pll0}
& P_{\alpha\beta}(\tau) - \lambda_{\alpha}(\tau) \lambda_{\beta} (\tau)=0\; .
 \end{eqnarray} 
Here $\tau$ is a proper time parametrizing   
a worldline $W^1$ 
in generalized superspace, 
\begin{eqnarray}\label{W1} 
W^1 \in \Sigma^{({n(n+1)\over 2}|n)}: & \;   
X^{\alpha\beta}=X^{\alpha\beta}(\tau)  \, , \quad  
\theta^{\alpha}=\theta^{\alpha}(\tau)\, ,   
\end{eqnarray}
$X^{\alpha\beta}(\tau)$ and $\theta^{\alpha}(\tau)$ are bosonic 
and  fermionic coordinate functions.  

The most general supersymmetry algebra 
(called M--algebra in $D=11$ case, {\it i.e.} for $n=32$, \cite{M-alg}) 
\begin{eqnarray}\label{QQZ} 
    \{ Q_\alpha, Q_\beta\}=P_{\alpha\beta}\quad , 
   \quad [Q_\alpha, P_{\alpha\beta}]=0
    \; ,
\end{eqnarray}
is realized in the model of \cite{BL98} on the Poisson brackets 
(for simplicity, we ignore the $i$ factor appearing in the Poisson 
brackets). 
After quantization, {\sl schematically} (see \cite{BL98,BLS99} for 
a precise Hamiltonian analysis and quantization),  
Eq.  (\ref{Pll}) could be considered as 
a condition  on the state vector $|\lambda>$ of the quantum dynamical system, 
\begin{eqnarray}\label{BPS0} 
P_{\alpha\beta}|\lambda>= \lambda_{\alpha}\lambda_{\beta}|\lambda> \; .
\end{eqnarray}
Such a state was called {\sl BPS preon} in \cite{BPS01} 
for reasons that will become clear below. Eq.  (\ref{BPS0}) implies  
\begin{eqnarray}\label{BPS1} 
\{ Q_\alpha, Q_\beta \} |\lambda> = \lambda_\alpha \lambda_\beta |\lambda> \; .
\end{eqnarray}
Then, introducing an auxiliary set of $(n-1)$  contravariant $GL(n)$ vectors 
$w^{\alpha}_I$ ($SO(t,D-t)$ spinors)  orthogonal to 
the covariant $GL(n)$ vector $\lambda_\alpha$,  
\begin{eqnarray}\label{mul} 
w^{\alpha}_I \lambda_\alpha = 0\; , \qquad 
I=1, \ldots , (n-1)\, , 
\end{eqnarray}
one finds 
$\; w^{\alpha}_I 
\{ Q_\alpha, Q_\beta \} |\lambda> = 0 \;$. 
As a result, one can conclude that the BPS preon state $|\lambda>$ 
preserves all but one supersymmetries \cite{BPS01},  
\begin{eqnarray}\label{BPS3} 
Q_I |\lambda> \equiv w^{\alpha}_I Q_\alpha |\lambda> 
= 0 \; , \quad I=1, \ldots , (n-1)\, .
\end{eqnarray}
Let us stress that the set of $(n-1)$ vectors  $w^{\alpha}_I$ is pure 
auxiliary and has been introduced  for convenience only. 
The preservation of $(n-1)$ of the $n$ supersymmetries by the state 
$|\lambda>$ is encoded in the fact that the eigenvalue matrix 
$\lambda_\alpha \lambda_\beta$ 
of the operator $\{ Q_\alpha, Q_\beta \}$, Eq. (\ref{BPS1}), has rank one.

Note that the causal structure of the symplectic spacetime 
$\Sigma^{({n(n-1)\over 2}|0)}$ found in \cite{V01c}  
is related to the observation that the state $|\lambda >$ obeying 
Eq. (\ref{BPS0}) provides 
the {\sl general solution} 
of the conformal high--spin wave equation \cite{V01s}
\begin{eqnarray}\label{BPS4} 
(P_{\alpha\beta}P_{\gamma\delta} - P_{\alpha\gamma}P_{\beta\delta})
|\lambda>= 0 \; .
\end{eqnarray}

The algebra similar to (\ref{QQZ}) 
is satisfied by the fermionic constraints $D_\alpha (\tau)\; $ 
($\,\{ D_\alpha, Q_\beta\}= 0\,$), 
\begin{eqnarray}\label{DDP} 
    \{ D_\alpha, D_\beta\}= - P_{\alpha\beta}\quad , 
   \quad [D_\alpha, P_{\alpha\beta}]=0
    \; . 
\end{eqnarray} 
Eqs. (\ref{DDP}) and  (\ref{Pll0})
imply that $(n-1)$ of $n$ fermionic constraints, 
$D_I = w^{\alpha}_I D_\alpha$, 
 are of the first class. 
These first class constraints generate $(n-1)$ local fermionic 
$\kappa$-symmetries through the Poisson brackets. Thus the number of 
$\kappa$--symmetries of the worldline actions \cite{BL98} coincides 
with the number of supersymmetries preserved by a BPS preon state
(see \cite{BPS01}) as one might expect 
(see, {\it e.g.} \cite{BKO}, and  \cite{BdAI2} for extended discussion). 
Thus, 
one can consider {\sl  
the presence of $(n-1)$ $\; \kappa$--symmetries as 
the main characteristic property of a BPS preon model 
in a superspace with $n$ fermionic coordinates}.

The states  $|\lambda>$ 
 were used in \cite{BPS01} to provide a complete 
algebraic classification of the BPS states in M--theory 
(hence the name of BPS  preons). 
This suggests to conjecture 
that  any BPS state $|\Psi_k >$ 
is a superposition of a definite  number $k$   
of the BPS preons. 
The number $k$ is determined by the rank of its  
generalized momentum matrix $p_{\alpha\beta}$,  
\begin{eqnarray}\label{BPSk} 
P_{\alpha\beta}|\Psi_k >= 
p_{\alpha\beta}|\Psi_k >\; , \quad 
rank ( p_{\alpha\beta})= k \; .
\end{eqnarray}
Then (see \cite{BPS01}) there exists a set of $k$ $GL(n)$ vectors 
$\lambda^a_{\beta}$ (spinors of the Lorentz $SO(t,D-t)\subset GL(n)$)  
such that  
\begin{eqnarray}\label{BPSkp} 
& p_{\alpha\beta}= 
\sum\limits_{a=1}^{k} \lambda_{\alpha}^a\lambda^a_{\beta} \; . 
\end{eqnarray}
Eq. (\ref{BPSkp}) allows to speculate that the BPS state $|\Psi_k >$, 
satisfying Eq. (\ref{BPSk}),   
can be considered as composite of $k$ preon states 
$|\lambda^a>$, $a=1, \ldots ,
k$ \cite{BPS01}.

The existence of BPS preons and  other BPS states preserving more than 
$1/2$ of the supersymmetry (i.e. composites of  
$k< n/2$ preons) 
is allowed from an algebraic point of view 
\cite{BL98,Hull99,GGHT00}. However, for a long time  realizations 
of such states as solitonic solutions of the 'usual' $D\leq 11$ supergravity 
equations were not known and, in fact, 
the first search in simple models gave negative results \cite{GGHT00}. 
However, 
such solutions (now with up to $28$ of $32$ supersymmetries 
preserved) have recently been found 
\cite{CLP02,GH02,BJM02} as a particular case of 
pp--waves \cite{PP}. Thus the original 
expectation that BPS preons and the states composed from 
less than $n/2$ BPS preons cannot be realized in the 'usual' superspace  
(a `BPS preon conspiracy')  is  broken, at least partially.  
The relation of such solutions with models in generalized superspace 
has not been clarified yet. One may assume that  
the (constant) 'values' of antisymmetric tensor fields, characteristic
of the $pp$--wave background, should play there the role of 
some tensorial coordinates of generalized superspace 
({\it cf.} \cite{JdA00}), but the details of the 
embedding of pp--wave spacetimes into a generalized superspace require 
additional study. 

Here we address another problem. 
Only point--like models 
with the properties of BPS preons \cite{BL98} (and composites of less than 
$n/2$ preons \cite{BL98'}) were known in the  generalized superspace. 
On the other hand, if one takes seriously the hypothesis 
\cite{BPS01} that all the M-theory BPS--states ($M2$--brane, 
$M5$--brane, intersecting brane configurations, {\it etc.}) are composed 
from ($n=32$) BPS preons, 
one should find for the latter an extended object  
model ({\it i.e.} a model with $p$--dimensional worldvolume 
$W^{p+1}$ rather than worldline $W^1$ (\ref{W1})), at least in 
the generalized superspace. 
The main message of this letter is that such a model for 
$D=11$ BPS preons is provided by a 'twistor--like' formulation of 
tensionless p--branes in the generalized superspace $\Sigma^{(528|32)}$. 
Moreover, the model can be formulated in an arbitrary generalized superspace 
 $\Sigma^{({n(n+1)\over 2}|n)}$.

Tensionless p--branes in $D=4$ ($n=4$) generalized superspace 
 $\Sigma^{(10|4)}$ were previously  
studied in \cite{ZL,ZU}. In \cite{ZU} it was found that a  
twistor--like formulation of the tensionless p--brane   
in $\Sigma^{(10|4)}$ (which generalizes the model from \cite{BZ} 
for the case of additional tensorial coordinates) 
possess 3 $\kappa$--symmetries. 
We will show here that for any $n$ (or any $D$), including $n=32$ ($D=11$), 
the $\Sigma^{({n(n+1)\over 2}|n)}$ generalization of the tensionless p--brane 
action from \cite{BZ} possesses $(n-1)$ $\kappa$--symmetries. 
In the light of the above mentioned correspondence, this implies that  
the $n= 32$ ($D=11$) version of this action 
provides a dynamical model for a BPS state which preserves  
$31$ of $32$ supersymmetries, {\it i.e.} an extended object model for 
a BPS preon.

\section{Tensionless $p$--brane action in $\Sigma^{({n(n+1)\over 2}|n)}$}
 
We consider the following 
action for an extended object ($p$--brane) moving in  generalized 
superspace $\Sigma^{({n(n+1)\over 2}|n)}$   
\begin{eqnarray}\label{Snull} 
S = \int d^{p+1}\xi \, L 
= {1\over 2} \int d^{p+1}\xi \, \rho^m \Pi_m^{\alpha\beta} 
\lambda_{\alpha}\lambda_{\beta} \; 
\end{eqnarray}
({\it cf.} \cite{BZ} for the usual $D=4$ superspace and \cite{B90} 
for $p=0$). 
Here 
\begin{eqnarray}\label{Pi} 
\Pi^{\alpha\beta} \equiv d\xi^m \Pi_m^{\alpha\beta} = 
dX^{\alpha\beta}(\xi) - i d\theta^{(\alpha}\,\theta^{\beta )}(\xi) 
\end{eqnarray}
is the pull--back of the supersymmetric Volkov--Akulov one--form for 
$\Sigma^{({n(n+1)\over 2}|n)}$ on the worldvolume   
\begin{eqnarray}\label{Wp} 
W^{p+1} \subset \Sigma^{({n(n+1)\over 2}|n)}: & \qquad 
X^{\alpha\beta}=X^{\alpha\beta}(\xi)\, , \; \qquad 
\theta^{\alpha}=\theta^{\alpha}(\xi)
\end{eqnarray} 
parametrized by local coordinates $\xi^m$, $\; m=0,1,\ldots , p$; 
$\rho^m= \rho^m(\xi)$ is a Lagrange multiplier and 
$\lambda_{\alpha}= \lambda_{\alpha}(\xi)$ are auxiliary bosonic 
variables. 
The action (\ref{Snull}) does not contain any dimensionful parameter, 
what  allows us to call its associated dynamical system 
{\sl tensionless super--p--brane}  
{\sl in generalized superspace}.

The $n=4$ counterpart of the action 
(\ref{Snull}), with $\lambda$ treated as a Majorana representation of  
$D=4$ Lorentz harmonics \cite{B90}, was studied in \cite{ZU}. 
On the other hand, for $n=2^k=dim(Spin(1,D-1))$, 
substituting $\Gamma_\mu^{\alpha\beta}\Pi_m^\mu
\equiv \Gamma_\mu^{\alpha\beta} (\partial_mx^\mu - i  
\partial_m\theta \Gamma^\mu \theta)$ for $\Pi_m^{\alpha\beta}$ in  
Eq. (\ref{Snull}), one arrives at a $D$ dimensional generalization of 
the null--super--$p$--brane action from \cite{BZ}. Certainly, only 
for $D=3,4,6,10$ the momentum density $P_\mu(\xi)= 
\lambda \Gamma_\mu \lambda$ is light--like 
and the tensionless super--$p$--brane can be called  null--super--p--brane. 

The set of global symmetries of the action  (\ref{Snull}) includes 
$GL(n)$ transformations acting on the indices $\alpha, \beta = 1,\ldots , n$. 
It is also invariant, by construction, under the global supersymmetry 
\begin{eqnarray}\label{susy} 
\delta_{\epsilon} X^{\alpha\beta}(\xi) = i\epsilon^{(\alpha}\theta^{\beta )}(\xi)\, , \quad \delta_{\epsilon} \theta^{\alpha}(\xi)= \epsilon^{\alpha}
\, , \quad \\ \nonumber 
\delta_{\epsilon} \lambda_{\alpha}(\xi)=0\; , \quad 
\delta_{\epsilon} \rho^m (\xi)=0\; , 
\end{eqnarray} 
The generators $Q_\alpha$ of the supersymmetry (\ref{susy}) 
 satisfy the algebra (\ref{QQZ}) involving the generator $P_{\alpha\beta}$  
of the translations: $\delta_{a} X^{\alpha\beta}(\xi )= a^{\alpha\beta}$, 
$\delta_{a} \theta^{\alpha}(\xi)=0$, $\delta_{a} \lambda_{\alpha}(\xi)=0$, 
$\delta_{a} \rho^m (\xi)=0$.

A straightforward calculation of canonical momentum for 
$X^{\alpha\beta}(\tau)$, 
$\; P_{\alpha\beta}= {\partial L \over \partial_0 X^{\alpha\beta}}$, 
results in the primary constraint 
\begin{eqnarray}\label{Pllp} 
\Phi_{\alpha\beta}= P_{\alpha\beta}(\xi)  - 
\rho^0 (\xi) \lambda_{\alpha}(\xi)\lambda_{\beta}(\xi)= 0 \; ,  
\end{eqnarray}
({\it cf.} Eq. (\ref{Pll0}))
which implies the propagation of the extended object in the 
directions characterized by 
$\lambda_{\alpha}(\xi)$. Such directions could be regarded as a 
$\Sigma^{({n(n+1)\over 2}|n)}$ 
generalization of the light--like directions of the usual $D$--dimensional 
superspace. 

The calculation of the other canonical momenta, ${\cal P}^\alpha (\xi) =  
{\partial L \over \partial (\partial_0 \lambda_{\alpha})}$, 
$\; {\cal P}_m =  {\partial L \over \partial (\partial_0 \rho_m)}$ and 
$\pi_\alpha (\xi) =  {\partial L \over \partial_0 \theta^{\alpha}}$ also 
results in the constraints: ${\cal P}^\alpha (\xi) =0$, 
$\; {\cal P}_m =0$ and 
\begin{eqnarray}\label{Dal} 
D_{\alpha}= \pi_{\alpha\beta}(\xi)  + i P_{\alpha\beta}
\theta^{\beta}(\xi)= 0 \; .   
\end{eqnarray}
The fermionic constraints (\ref{Dal})  
 obey the algebra (\ref{DDP}) on the Poisson brackets. 
This already indicates the presence of 
$(n-1)$ local fermionic $\kappa$--symmetries, which we now 
describe explicitly in the Lagrangian approach.

\section{$\kappa$--symmetry and other gauge symmetries} 
It  
is convenient to write the general variation 
of the action (\ref{Snull}) as   
\begin{eqnarray}\label{vSnull} 
& \delta S & = 
\int d^{p+1}\xi \, [ {1\over 2} \delta \rho^m \Pi_m^{\alpha\beta} 
\lambda_{\alpha}\lambda_{\beta}
+ \rho^m \Pi_m^{\alpha\beta} 
\lambda_{\beta} \, \delta  \lambda_{\alpha}] - 
\nonumber  \\ 
& & - {1\over 2}
\int d^{p+1}\xi \; \partial_m (\rho^m
\lambda_{\alpha}\lambda_{\beta})  i_\delta \Pi^{\alpha\beta} -
  i \int d^{p+1}\xi \; \rho^m  \partial_m \theta^{\alpha} 
\lambda_{\alpha}  \delta \theta^{\beta} 
\lambda_{\beta} \; ,   
\end{eqnarray}
where 
$i_\delta \Pi^{\alpha\beta} \equiv \delta X^{\alpha\beta} - 
i \delta\theta^{(\alpha} \, \theta^{\beta )}\,$,    
and integration by parts has been performed. 
Eq. (\ref{vSnull}) makes evident that  the action (\ref{Snull}) possesses 
{\sl $(n-1)$ $\kappa$--symmetries} 
\begin{eqnarray}
\label{kappaL} 
\delta_{\kappa} \rho^m = 0 \; , \qquad \delta_{\kappa} \lambda_{\alpha}=0 \; ,
& \\ 
\label{kappaX} 
\delta_{\kappa} X^{\alpha\beta}(\xi) = 
i\delta_{\kappa}\theta^{(\alpha}\theta^{\beta )}(\xi)\; , 
\quad &
\\ \label{kappaTh} 
\delta_{\kappa} \theta^{\alpha}(\xi)\,= \kappa^I(\xi) w_I^{\alpha}(\xi) 
\; , \qquad & I=1,\ldots , (n-1)\;  ,  
\end{eqnarray} 
with parameters $\kappa^I(\xi)$. 
In Eq. (\ref{kappaTh}) the $(n-1)$ auxiliary  
$GL(n)$ vector fields  $w_I^{\alpha}(\xi)$ are defined as in  Eq. (\ref{mul}), 
$w_I^{\alpha}(\xi) \lambda_{\alpha}(\xi)=0$.
In other words, 
the $\kappa$--symmetry transformation of 
the Grassmann  coordinate function (\ref{kappaTh}) is provided by the 
general solution of the equation
\begin{eqnarray}
\label{kappaTh0} 
\delta_{\kappa} \theta^{\alpha}(\xi)\, \lambda_{\alpha}(\xi) =0
\; . \qquad 
\end{eqnarray}
Thus, we are not enforced to consider an extension of the phase space of our 
dynamical system by incorporation of auxiliary variables 
$w^{\alpha}_I(\xi)$ and their momentum: we     
can keep Eqs. (\ref{kappaX}), (\ref{kappaL}), 
(\ref{kappaTh0}) instead as the definition of the $\kappa$--symmetry 
(but we may use $w^{\alpha}_I(\xi)$ as a convenient tool 
to present the results in  a transparent form).

The bosonic 'superpartner' of the fermionic $\kappa$--symmetry 
is provided by the b--symmetry transformations   
\begin{eqnarray}
& \delta_{b} \theta^{\alpha}(\xi)\,= 0 \; , \qquad  \delta_{b} \rho^m = 0\;, 
 \qquad  \delta_{b} \lambda_{\alpha}=0 \; 
\nonumber
\\ 
\label{bX}  
& 
\delta_{b} X^{\alpha\beta}(\xi) = b^{IJ}(\xi) 
w_I^{\alpha}(\xi) w_J^{\beta}(\xi)
\; ,   
\quad 
\end{eqnarray} 
with ${n(n-1)\over 2}$ parameters $b^{IJ}(\xi)=  b^{JI}(\xi)$, 
$\; I, J=1,\ldots , (n-1)$.  
The only nontrivial part of the $b$--symmetry transformations, Eq. (\ref{bX}), 
is  the general solution of the equation 
\begin{eqnarray}
\label{bX0}  
\delta_{b} X^{\alpha\beta}(\xi) \,\lambda_{\beta}(\xi) = 0 \; . 
\quad 
\end{eqnarray} 
Note also an evident scaling gauge symmetry of the action
(\ref{Snull}),  
\begin{eqnarray}
 \delta_{s} \theta^{\alpha}(\xi)\,= 0\; , \qquad 
 \delta_{s} X^{\alpha\beta}(\xi) = 0\; , 
\nonumber 
\\ 
\label{slb}  
\delta_{b} \rho^m = - 2 s(\xi)  \rho^m \; , \qquad 
\delta_{s} \lambda_{\alpha}= s(\xi)\lambda_{\alpha}(\xi)
\; ,
\end{eqnarray} 
as well as the symmetry under worldvolume general coordinate transformations 
(in their variational version $\tilde{\delta}_{g.c.}$ 
characterized by $\tilde{\delta}_{g.c.}\xi^m=0$, see  
\cite{BdAI2} and refs. therein)
\begin{eqnarray}\label{gcX}  
 & \tilde{\delta}_{gc} X^{\alpha\beta}(\xi) 
= t^m(\xi) \partial_mX^{\alpha\beta} 
\; , & 
\tilde{\delta}_{gc}\theta^{\alpha}(\xi)  = t^m(\xi) \partial_m \theta^{\alpha} 
\; ,  \quad  
\tilde{\delta}_{gc}\lambda_{\alpha}(\xi) = t^m(\xi) \partial_m \lambda_{\alpha}
\; , 
\\ \label{gcr}
& & \tilde{\delta}_{gc} \rho^m (\xi) = 
\partial_n (\rho^m t^n) - \rho^n \partial_n t^m \; . 
\qquad 
\end{eqnarray}

\section{ Supertwistor representation and 
$OSp(64|1)$ symmetry of the BPS preon model}   

Let us use  the Leibniz rule 
($\lambda \partial_m X\equiv 
\partial_m (\lambda X)- 
(\partial_m \lambda) X$, {\it etc.},  
no integration by parts and no gauge fixing) to present the action (\ref{Snull}) in 
the equivalent form
\begin{eqnarray}\label{Snull1} 
S = &
 {1\over 2} \int d^{p+1}\xi \, 
(\lambda_{\alpha} \rho^m \partial_m \mu^\alpha - 
\rho^m \partial_m \lambda_{\alpha}\; \mu^\alpha) - 
\\ \nonumber & - {i\over 2} \int d^{p+1}\xi \,  
\rho^m \partial_m \eta \; \eta )
\; ,
\end{eqnarray}
where 
\begin{eqnarray}
\label{mu} 
& \mu^\alpha = X^{\alpha\beta} \lambda_{\beta}- {i\over 2} 
\theta^{\alpha} \theta^{\beta} \lambda_{\beta} \; , \quad 
\eta = \theta^{\beta} \lambda_{\beta} \; .
\end{eqnarray}
$\lambda_{\alpha}$,  
$\mu^\alpha$ and $\eta$  
can be regarded as components of an $OSp(2n|1)$ supertwistor ${\cal Y}^\Sigma$ 
\cite{BL98},
\begin{eqnarray}
\label{Y} 
{\cal Y}^\Sigma = \left(\matrix{\lambda_{\alpha} \, , & 
\mu^\alpha \, , & \eta }\right) \; 
\end{eqnarray}
In terms of ${\cal Y}^\Sigma$ the action (\ref{Snull1}) reads 
\begin{eqnarray}\label{Snull2} 
S =  - {1\over 2} \int d^{p+1}\xi \, \rho^m \partial_m {\cal Y}^\Sigma 
\; C_{\Sigma\Lambda} \; {\cal Y}^\Lambda \; ,
\end{eqnarray}
where 
\begin{eqnarray}\label{Csym} 
C_{\Sigma\Lambda} = \left(\matrix{ 0 & \delta^{\alpha}{}_{\beta} & 0  
\cr 
- \delta_{\alpha}{}^{\beta} & 0 & 0 \cr 
 0 & 0 & i \cr }\right) = - (-)^{\Sigma \Lambda} C_{\Lambda\Sigma}
\end{eqnarray}
is the orthosymplectic ($OSp(2n|1)$ invariant) 'metric' tensor. 

The Lagrange multiplier $\rho^m$ does not carry physical degrees of freedom. 
Indeed, using the general coordinate transformations $\tilde{\delta}_{gc}$, 
Eq. (\ref{gcr}), and the scaling symmetry, Eq. (\ref{slb}), 
one can fix, {\it e.g.}, the gauge $\rho^m(\xi)= \delta^m_0$.  
 The generalized Penrose correspondence (\ref{mu}) clearly does not restrict 
$\mu^\alpha$ (as the first term in {\it r.h.s} contains the ${n(n+1)\over 2}$ 
parametric 
$X^{\alpha\beta}$). 
Hence the tensionless super--$p$--brane model allows a description 
in terms of $2n$ bosonic and $1$ fermionic components 
of the unconstrained {\sl orthosymplectic supertwistor 
(\ref{Y})} which {\sl describes all the physical degrees of 
freedom of the system} and makes 
the global $OSp(2n|1)$ symmetry manifest. 
In particular, this implies that for $n=32$
(i.e. $D=11$) 
the extended BPS preon model (\ref{Snull}) possesses an $OSp(64|1)$ 
generalized superconformal symmetry, 
which is characteristic both for high--spin 
theories (see \cite{V01,V01s,V01c})  and for the two--time physics 
approach to M-theory \cite{Bars2t,West}.

\section{Conclusion and outlook}

We have shown that the dynamical system described  by the action (\ref{Snull}) 
 possesses $(n-1)$ local fermionic 
$\kappa$--symmetries. Hence, in $n=32$ ($D=11$) such a dynamical system  
 can be considered as 
an extended object model for BPS preons, the hypothetical constituents 
of M-theory \cite{BPS01}. 
We have seen as well that the BPS preon model possesses  
$OSp(64|1)$ symmetry, 
which was suggested to be a generalized conformal symmetry of M-theory 
(see \cite{Bars2t,West,BPS01} and refs. therein);  
this becomes transparent after passing to the equivalent  
supertwistor representation, Eq. (\ref{Snull1}) or (\ref{Snull2}), 
of the action (\ref{Snull2}). This simple transformation  also exhibits 
the physical  degrees of freedom of the dynamical system.

We call the object described by the action 
(\ref{Snull}) a 
{\sl tensionless super--p--brane in generalized superspace} 
 $\Sigma^{({n(n+1)\over 2}|n)}= 
\{ (X^{\alpha\beta}, \theta^{\alpha})\}$. The reasons are that  
the  action (\ref{Snull}) does not contain dimensionful parameters, and 
that the constraints (\ref{Pllp}) imply propagation in the 
generalized light--like directions of $\Sigma^{({n(n+1)\over 2}|n)}$ 
({\it cf.} Ref. \cite{V01s}). 
Moreover, for $n=2,4,8,16$, 
one converts Eq. (\ref{Snull}) into the action of null--super--p--brane 
in the usual $D=3, 4, 6, 10$ superspaces (see \cite{BZ} for D=4) by 
substituting $\Gamma_\mu^{\alpha\beta}\Pi_m^\mu
\equiv \Gamma_\mu^{\alpha\beta} (\partial_mx^\mu - i 
\partial_m\theta \Gamma^\mu \theta)$ for $\Pi_m^{\alpha\beta}$.

Tensionless strings and $p$--branes in usual spacetime 
and usual superspace  
were discussed many times in the context of superstring/M--theory 
\cite{Shild}--\cite{Zh}, \cite{BZ}, \cite{Lindstrom2}--\cite{CEGK}
(see \cite{BZ,Bozhilov1} for more references). In particular, 
they appear as singularities in K3 compactification of superstring theory 
down  to six dimensions which connect all known supersymmetric six 
dimensional vacua \cite{Witten}. 
An interesting perturbative approach to  search for solutions  
of nonlinear superstring equations in the curved spacetime background 
was developed in \cite{Zh}. It is based on a power series decomposition 
in the $p$--brane tension  $T_p$ and 
is close in spirit to earlier propositions \cite{Isham} 
to obtain the quantum propagator of a $p$--brane by starting from the 
propagator 
of null--$p$--brane and summing up the perturbative series in $T_p$. 
The leading order of such expansion, null--string for $p=1$, 
should dominate string amplitudes describing short distance string  physics 
\cite{Gross}.  
The tension generation mechanism, which allows one to obtain 
a tensionful superbrane action from a null--super--p--brane action 
was studied in \cite{ten-gen,ten-gen1}. In this frame 
the (super)brane tension $T_p$ appears as an integration 
constant in the solution of the superstring equations of motion. 
This allows for its different values in  regions of a  
universe separated by a domain wall and unifies null--$p$--branes 
($T_p=0$) with tensionful $p$--branes ($T_p\not= 0)$ . 
Furthermore, it was shown in \cite{ten-gen1} that the tension 
$T_p$ can appear also as a result of a dimensional reduction. 
A development of this approach for the case of generalized superspaces 
might be useful for establishing mechanisms of tension generation and 
of the formation of the fundamental M--branes from our extended BPS preons.

Suggestions about a possible relation between tensionless strings and 
higher spin field theories can be found already in 
\cite{Vasiliev89}. Recently this possible relation was discussed 
in the context of AdS/CFT correspondence \cite{Sundborg}. The key 
observation is that, on one hand, both string field theory 
and the interacting higher spin theory contain infinite number 
of fields of higher spins, but, on the other hand,  
the latter has much more powerful gauge symmetry. This allowed Vasiliev to 
discuss in \cite{Vasiliev89} the possibility  
that higher spin theories are more 
fundamental than string theory and that  string theory can be viewed 
as a spontaneously broken phase of the higher spin theories. 
Then a possibility of an identification of 
the null strings (or null--$p$--branes) 
with higher spin theories was suggested by the enhancement of the symmetry 
in the tensionless limit of (super)string model 
(see \cite{Sundborg} for further reasons). 

The BPS preon conjecture \cite{BPS01} as a whole and, particularly,  
the tensionless superstring and 
super--p--brane models (\ref{Snull}) {\sl in generalized 
superspace}, can be considered also as a development of the above ideas. 
The models (\ref{Snull}), being formulated in the generalized superspace 
which allows for a formulation of higher spin theories \cite{Fr86,V01s,V01c},  
respect, by construction, at least the $GL(n)$ part of 
the higher spin symmetry 
(see \cite{ZL,ZU} for other models in $D=4$). 
The physically relevant M--branes and D--branes 
in usual $D=11$ and $D=10$ superspaces are expected to 
appear in a spontaneously broken phase
of the BPS preon models, which should imply the breaking 
of $GL(n)$ symmetry down to some $Spin(1,D-1) \subset GL(n)$
(e.g. $Spin(1,10) \subset GL(32)$ for M-branes, 
$Spin(1,9) \subset GL(32)$ for D--branes). 
Moreover, for $n=2,4,8,16$ the models (\ref{Snull}) are directly related 
the $D=3,4,6, 10$ massless  higher spin theories: for $p>0$  they 
describe an extended object generalization of the 
classical mechanics description of 
the free higher spin theories. 
Indeed, as it was shown in 
\cite{BLS99}, 
 the quantum state spectra  of the $n=2,4,8,16$ 
generalized superparticle models \cite{BL98}, which 
are identical to the 
point--like ($p=0$)  
models (\ref{Snull}) 
($\rho^0$ can be removed by rescaling of  $\lambda$, Eq. (\ref{slb})), 
consist of 
towers of massless fields of all possible `spins' in 
$D=3,4,6\; and \; 10$. 
The special property of the  
$n=2,4,8,16$ ($n=2(D-2)$, $D=3,4,6, 10$) point--like 
models (\ref{Snull}) are related with the existence of the Hopf fibrations 
$S^{2D-5}/S^{D-2}= S^{D-3}$ (see \cite{BLS99}). 
This provides a mechanism for `momentum space compactification' of 
the additional (with respect to usual spacetime) degrees of freedom. 
The situation with 
$n=32$ ($D=11$) model is still unclear: 
classically it describes a particle with a dynamically generated mass 
\cite{BL98,BL98'} and there is a problem in interpretation of the 
quantum state spectrum because no counterpart of the Hopf fibration 
is known for this case (see \cite{BL98'} for some discussion). 
However, in the framework of the BPS preon conjecture \cite{BPS01}, 
which refers to superbrane rather than to field theories, 
the problem is rather a search for possibilities  to construct a  
`physical' BPS objects defined in the standard superspace, 
like M--branes and D--branes, from  BPS preons in generalized superspace. 
In principle, one could explore the composite nature of the M--branes in terms
of the    point--like BPS preon model,  but in an indirect way 
similar to the Matrix model description of supermembrane  \cite{MT}. 
The tensionless $p$--brane models (\ref{Snull}) provide a new basis to 
search for a possible composed nature of the M--branes: 
this search might be carried out by the quasi-classical methods 
for the extended object action, {\it e.g.} by studying solutions of 
the  equations of motion (see \cite{ZL} for some results in $D=4$) 
and specific interactions with background fields in generalized superspace.

Recently an explicit relation between superparticle wavefunctions 
in generalized superspace and  Vasiliev's 'unfolded equations' 
for higher spin field was established in \cite{Dima} for $n=4$ ($D=4$). 
Moreover, the quantization  
of a counterpart of the $p=0$, $n=4$ model (\ref{Snull})  
defined in the generalized $AdS_4$ superspace 
has been also considered in \cite{Dima}. 
This is of particular importance as the nontrivial interactions 
of higher spin fields can be constructed in a selfconsistent way 
only in a spacetime with nonvanishing cosmological constant 
(see \cite{Vasiliev89} and refs. therein).
It is interesting that the proper generalized $AdS_4$ superspace 
was proved to be just the supergroup manifold $OSp(1|4)$ \cite{Dima,Misha}. 
These results provide a reason to study also 
the $AdS$ generalizations of the $n=4$ ($D=4$) 
versions of the model (\ref{Snull}): the 
tensionless superstring and supermembrane on  $OSp(1|4)$ supergroup manifold.

\bigskip

{\it Acknowledgments}. 
The author  is grateful to J.A. de Azc\'arraga and to I. Bars, 
J. Lukierski, M. Vasiliev for useful  
conversations, and to  
the Abdus Salam ICTP for kind hospitality 
in Trieste.  This work has been partially supported by 
the research grant BFM2002-03681 from the Ministerio de Ciencia y 
Tecnologia and from EU FEDER, Ukrainian FFR grant 
$\# 383$ and INTAS grant N 2000-254.

\end{document}